# The Temple University Hospital Seizure Detection Corpus

*Vinit Shah, Eva von Weltin, Silvia Lopez, James Riley McHugh,*
*Lily Veloso, Meysam Golmohammadi, Iyad Obeid and Joseph Picone*

*Department of Electrical and Computer Engineering, Temple University, Philadelphia, PA, USA*

**Correspondence:**
**Vinit Shah**
**vinitshah@temple.edu**

## Introduction

The electroencephalogram (EEG), which has been in clinical use for over *70* years, is still an essential tool for diagnosis of neural functioning (Kennett, 2012). Well-known applications of EEGs include identification of epilepsy and epileptic seizures, anoxic and hypoxic damage to the brain, and identification of neural disorders such as hemorrhagic stroke, ischemia and toxic metabolic encephalopathy (Drury, 1988). More recently there has been interest in diagnosing Alzheimer's (Tsolaki et al., 2014), head trauma (Rapp et al., 2015) and sleep disorders (Younes, 2017). Many of these clinical applications now involve the collection of large amounts of data (e.g., 72-hour continuous EEG recordings), which makes manual interpretation challenging. Similarly, the increased use of EEGs in critical care has created a significant demand for high-performance automatic interpretation software (e.g., real-time seizure detection).

A critical obstacle in the development of machine learning (ML) technology for these applications is the lack of big data resources to support training of complex deep learning systems. One of the most popular transcribed seizure databases available to the research community, the CHB-MIT Corpus (Goldberger et al., 2000), only consists of *23* subjects. Though high performance has been achieved on this corpus (Shoeb, 2010), these results have not been representative of clinical performance (Golmohammadi, et al., 2018). Therefore, we introduce the TUH EEG Seizure Corpus (TUSZ), which is the largest open source corpus of its type, and represents an accurate characterization of clinical conditions.

Since seizures occur only a small fraction of the time in this type of data, and manual annotation of such low-yield data would be prohibitively expensive and unproductive, we developed a triage process for locating seizure recordings. We automatically selected data from the much larger TUH EEG Corpus (Obeid and Picone, 2016) that met certain selection criteria. Three approaches were used to identify files with a high probability that a seizure event occurred: (1) keyword search of EEG reports for sessions that were likely to contain seizures (e.g., reports containing phrases such as "seizure begins with" and "evolution"), (2) automatic detection of seizure events using commercially available software (Persyst, 2017), and (3) automatic detection using an experimental deep learning system (Golmohammadi et al., 2018). Data for which approaches (2) and (3) were in agreement were given highest priority.

Accurate annotation of an EEG requires extensive training. For this reason, manual annotation of EEGs is usually done by board-certified neurologists with many years of post-medical school training.

Consequently, it is difficult to transcribe large amounts of data because such expertise is in short supply and is most often focused on clinical practice. Previous attempts to employ panels of experts or use crowdsourcing strategies were not productive. However, we have demonstrated that a viable alternative is to use a team of highly trained undergraduates at the Neural Engineering Data Consortium (NEDC) at Temple University. These students have been trained to transcribe data for seizure events (e.g. start/stop times; seizure type) at accuracy levels that rival expert neurologists at a fraction of the cost (Shah et al., 2018). In order to validate the team's work, a portion of their annotations were compared to those of expert neurologists and shown to have an extremely high inter-rater agreement.

In this paper, we describe the techniques used to develop TUSZ, evaluate their effectiveness, and present some descriptive statistics on the resulting corpus.

**Method**

To build an annotated seizure dataset, we first needed an abundant source of EEG data. Our work here utilized a subset which includes approximately *90%* of v0.6.0 of TUH EEG. The data is organized by patient and by session. Each session contains EEG signal data stored in a standard European Data Format (EDF) (Kemp, 2013) and a de-identified report written by a board-certified neurologist. The EDF files contain a variable number of channels (Obeid & Picone, 2016) but during the annotation process only 19 EEG channels plus two supplementary channels (heart rate and photic stimulation) were used. The data were annotated using our open source annotation tool (Capp et al., 2017).

Since less than *0.1%* of the original data contains actual seizure events, annotating the entire database would be costly and inefficient. Therefore, we used three independent approaches to identify sessions that were likely to contain actual seizure events. First, we applied natural language processing (NLP) techniques to identify reports that had keywords related to ictal patterns. The reports were preprocessed using filters that normalized (e.g., removed punctuation and misspellings) and stemmed the text (Sirsat et al., 2013). Machine learning experiments were conducted that utilized term frequency-inverse document frequency (tf-idf) features (Manning et al., 2008). Popular machine learning approaches such as NegEx (Chapman et al., 2001), Naïve Bayes and Support Vector Machines with linear kernel functions (SVM) (Vapnik, 1995) were trained to recognize documents with that were most likely to contain seizure terms. The Naïve Bayes and Support Vector Machines algorithms used tf-idf features while the NegEx algorithm used raw features (e.g., words) for classification of reports as ictal or non-ictal. These algorithms were seeded from *197* reports describing the occurrence of a seizure and *2,471* reports describing non-occurrence of a seizure. All three algorithms were tested on *100* reports (*50* ictal and *50* non-ictal), with NegEx performing slightly better than the Naïve Bayes and SVM classifiers.

The classification of reports using NegEx was performed using a regular expression rule-based approach. The regular expression labels were selected based on negation (NEG), context (CNTX) and affirmation (AFFR). The negation labels were selected based on three different types of negations: pre-negation (PREN) (i.e. did not experience), post-negation (POST) (i.e. infiltrates were not shown) and pseudo-negation (PSEU). NegEx correctly classified *99%* of the reports used in our pilot study of *100* reports. When applied to *18,000* sessions in TUH EEG, *844* sessions were identified as likely to have a seizure. Of these *844* sessions, manual annotation determined that *174* sessions had actual seizures.

The second method used to triage the data was to process the data through a state of the art commercial software tool, P13 rev. B from Persyst Development Corporation (*https://www.persyst.com/technology/seizure-detection/*) (Persyst, 2017). We determined that *1,388* files out of *34,698* files contained seizure events. Our third method used an experimental seizure detection system known as AutoEEG (Golmohammadi et al., 2018). This system detected seizures with high confidence in *1,466* files out of *31,645* files. Files for which both systems agreed on a seizure were given the highest priority for annotation. These automated tools agreed on *146* files, or *0.42%*, of the corpus. The total number of sessions that were identified as having at least one seizure by either tool was *28*.

Using these three approaches, we identified *872* sessions containing *2,582* files from the original *16,168* sessions as high-yield data, meaning they were likely to contain seizures. Our annotation team then manually annotated all the data in these sessions and found that *280* of these sessions contained actual seizure events. It is interesting to note that of the three approaches for identifying high yield data, keyword search proved to be most effective. Automated seizure detection algorithms still suffer from poor performance, especially on short duration seizure events.

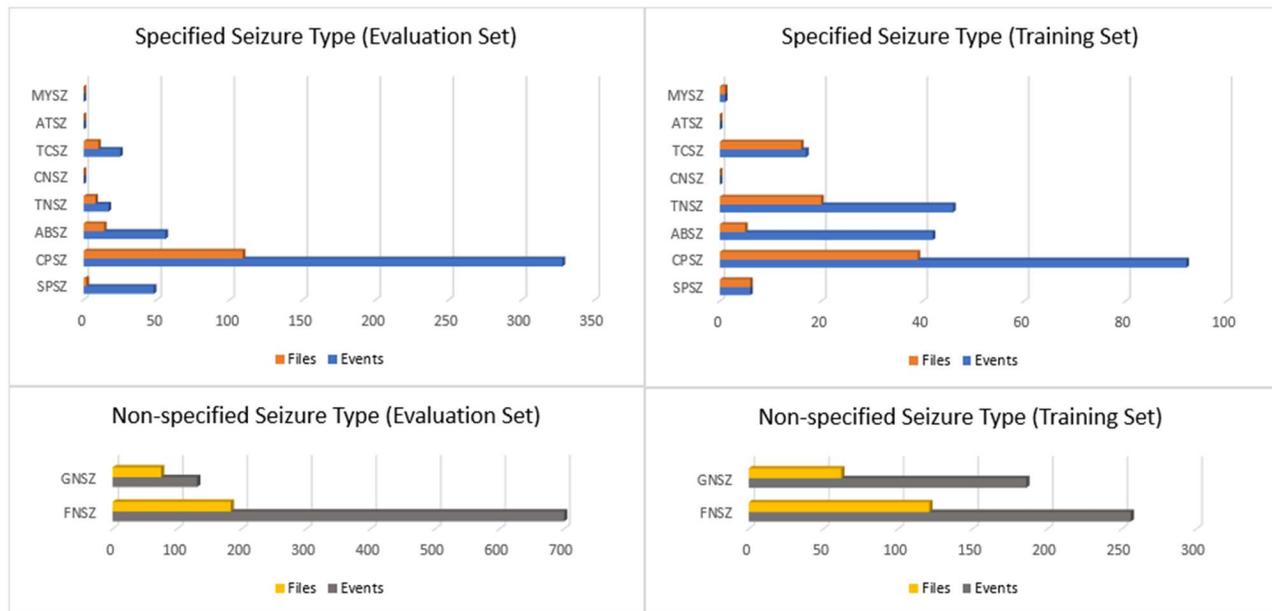

**Figure 1.** Histograms of seizure types in the TUH EEG Seizure Corpus for the evaluation and training sets.

**Results**

The most recent release of TUSZ is v1.2.0, which was released in December 2017. It contains *315* subjects with a total of *822* sessions, of which *280* sessions contain seizures. Each file is completely transcribed in two ways: channel-based and term-based. A channel-based annotation refers to labeling of the start and end time of an event on a specific channel. A term-based annotation refers to a summarization of the channel-based annotations – all channels share the same annotation, which is an aggregation of the per-channel annotations.

Based on the neurologist's report and careful examination of the signal, our annotation team was able to identify the type of seizures (e.g., absence, tonic-clonic). A list of these labels is shown below:

| | | | |
|---|---|---|---|
| SEIZ: | Seizure | | |
| GNSZ: | Generalized Non-Specific Seizure | TNSZ: | Tonic Seizure |
| FNSZ: | Focal Non-Specific Seizure | CNSZ: | Clonic Seizure |
| SPSZ: | Simple Partial Seizure | TCSZ: | Tonic Clonic Seizure |
| CPSZ: | Complex Partial Seizure | ATSZ: | Atonic Seizure |
| ABSZ: | Absence Seizure | MYSZ: | Myoclonic Seizure |

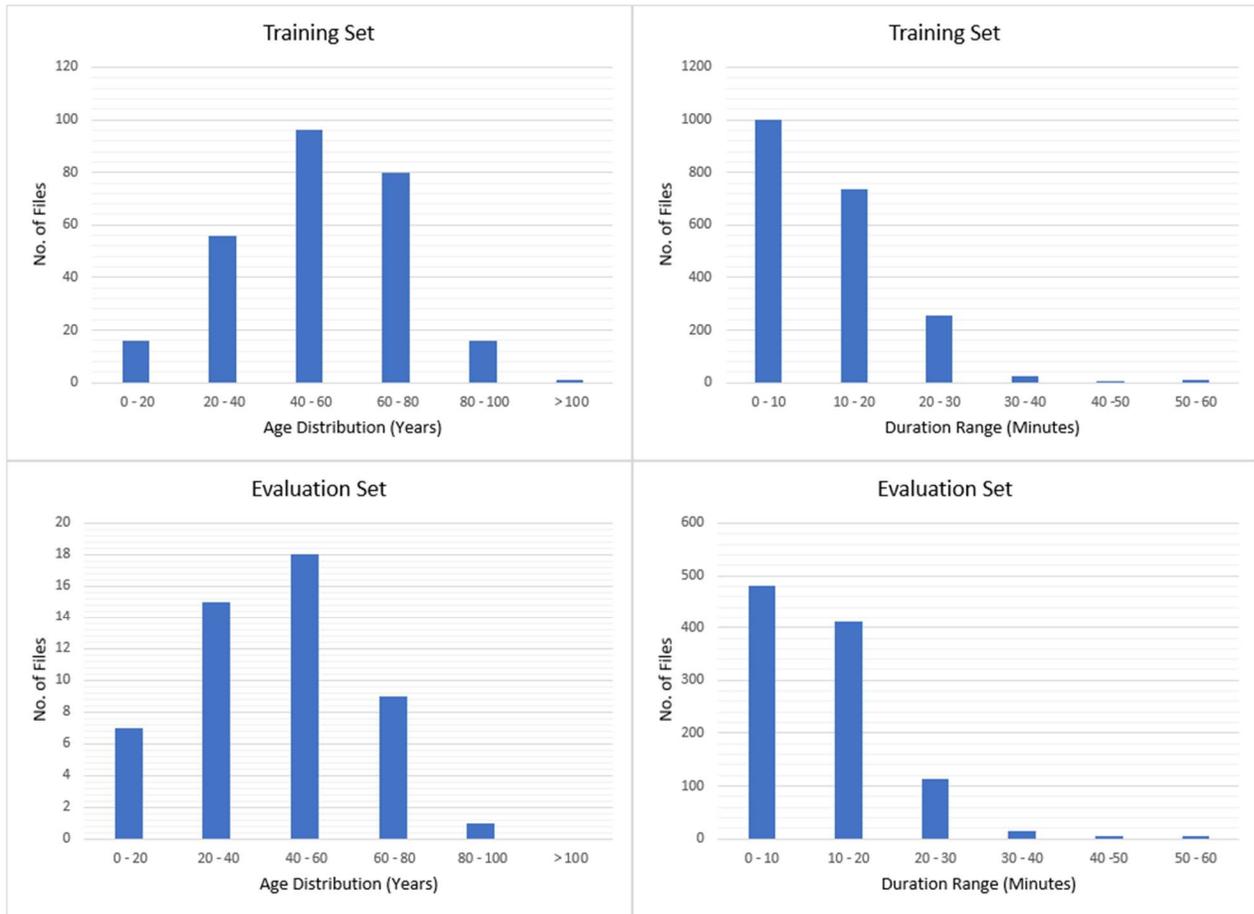

**Figure 2.** Histograms of age and duration.

If there was insufficient evidence to classify the type of seizure, then an event was defined as either "generalized non-specific" or "focal non-specific" depending on the focality. Histograms of the frequency of occurrence for these seizure types are shown in Figure 1.

We then segmented the data into a training and evaluation set to support technology development. The evaluation set was designed to provide a representative sampling of all conditions found in the training set under the constraint that it included *50* patients. Approximately *34%* of the evaluation dataset files contain seizures, which is much higher than typical clinical EEG data. The evaluation set was designed to be compact and yet provide representative results so that it would support rapid turnaround of experiments using a moderate amount of computational resources.

The entire seizure database has been divided into training and evaluation sets to support machine learning research. All files in this corpus are pruned versions of the original EEG recordings. The duration of a single pruned file is no more than one hour. The training and evaluation sets contain *265* and *50* subjects respectively. The patients in the evaluation set were selected based on gender (*56%* of the patients in the evaluation set are female; *50%* female in the training set) and selected to maximize a number of demographic features, as shown in Figure 2.

In addition to providing the raw signal data and annotations of seizure events, TUSZ contains metadata such as patient demographics, seizure type, and the type of EEG study. The EDF files contain the following metadata:

    patient id (anonymized)
    gender (male or female)
    age (measured in years due to privacy issues)
    recording data (DD-MMM-YYYY)
    per-channel information:
        labels, sample frequency, channel physical dimension, channel physical min, channel physical max,
        channel digital min, channel physical max, channel prefiltering conditions

We also have released a spreadsheet with the data that describes each patient and session in terms of the following fields:

    patient id (anonymized)
    session id
    EEG type / subtype:
        EMU / EMU (Epilepsy Monitoring Unit)
        ICU (Intensive Care Unit) /
            BURN (Burn Unit)
            CICU (Cardiac Intensive Care)
            ICU (Intensive Care Unit)
            NICU (Neuro-ICU Facility_
            NSICU (Neural Surgical ICU)
            PICU (Pediatric Intensive Care Unit)
            RICU (Respiratory Intensive Care Unit)
            SICU (Surgical Intensive Care Unit)
        Inpatient /
            ER (Emergency Room)
            OR (Operating Room)
            General
        Outpatient / Outpatient
        Unknown / Unknown (location cannot be determined)
    LTM or Routine
    Normal or Abnormal
    Number of Seizures per Session and File
    Start Time, Stop Time
    Seizure Type

The EEG Type and EEG Subtype fields are used to identify the general location of the EEG session with the hospital. A qualitative assessment of the duration of the recording is indicated a field that indicated whether the EEG was a routine recording (typically an outpatient session lasting 30 minutes) or an extended long-term monitoring (LTM). The normal/abnormal classification follows the clinical criteria described by Lopez (2017).

While most researchers can work with the information about seizure events provided in the above spreadsheet, we also provide a series of label files that allow display of seizure labels in a time-aligned manner using an open source visualization and annotation tool (Capp et al., 2017).

## Discussion

For deep learning technology to address problems such as seizure detection, large amounts of annotated data are needed. TUSZ is the world's largest publicly available corpus of annotated data for seizure detection that is unencumbered. No data sharing or IRB agreements are needed to access the data. The entire database consists of over *504* hours of data. Seizure events comprise about *36* hours or about *7%* of the data that has been annotated. Version 1.0.0 of the TUH EEG Corpus contains about 16,000 hours of data. We have not completed processing all of that data for seizure events, but our estimate is that the overall yield for seizure data using the process described in this paper is 0.2%. Since we are accessing pruned EEGs, the overall yield from continuous data is even smaller. This is a quite sobering statistic since it reveals the challenges in building the big data resources necessary to fuel deep learning research. Accurate triaging of the data is critical to building these resources in a cost-effective manner.

TUSZ contains a rich variety of seizure morphologies. Variation in onset and termination, frequency and amplitude, and locality and focality protect the evaluation and training sets against bias towards one type of seizure morphology. Models trained using this database will be strengthened by the mix of obvious and subtle seizure morphologies and will have the potential to be better prepared for applications handling real world data.

Seizures are a biological process that build gradually, often lacking discrete start and stop times. Event-based and term-based annotations are therefore included in our corpus in an effort to represent two different views of seizure evolution and duration. Event-based annotations are per-channel annotations and give users a very detailed account of where in the brain the seizure originates, how it spreads, and how it terminates while term-based annotations are the same on every channel and simply include the earliest seizure start time and the latest seizure end time. To address this issue, both multi-class and bi-class annotations have been made available. Multi-class annotations provide users more specific data on the type of seizure that is occurring, while bi-class annotations simply answer the question: is there a seizure occurring or not?

We are working continuously to improve expert knowledge of seizures that can be directly channeled into improving and expanding TUSZ. Development of annotation skills and increased use of automation will allow us to continue to improve the corpus. We have developed methods to automatically annotate other events, such as eye movements, generalized periodic discharges (GPD), periodic lateralized discharges (PLD), spikes, and sharp wave (Harati et al., 2015). We have also developed methods for cohort retrieval (Obeid et al., 2017) and parsing of EEG reports (Harabagiu et al., 2016). Though our focus is currently on seizure annotation, we will soon release more metadata related to TUSZ that will enable basic neuroscience research with the data.

TUSZ has been in beta release since late 2016 and can be downloaded from *https://www.isip.piconepress.com/projects/tuh_eeg/downloads/*. Users must register and provide a valid email address so that we can track usage. Users can also acquire the data by sending us a disk drive. Our rapidly growing userbase currently includes over *700* registered users.

## Acknowledgement

The authors would like to thank Dr. Mercedes Jacobson of Temple University Hospital and Dr. Steven Tobochnik of New York Presbyterian Hospital at Columbia University for enabling the creation of this corpus and their instruction in the interpretation of clinical EEG data.

## Funding

Research reported in this publication was most recently supported by the National Human Genome Research Institute of the National Institutes of Health under award number U01HG008468. The content is solely the responsibility of the authors and does not necessarily represent the official views of the National Institutes of Health. This material is also based in part upon work supported by the National Science Foundation under Grant No. IIP-1622765. Any opinions, findings, and conclusions or recommendations expressed in this material are those of the author(s) and do not necessarily reflect the views of the National Science Foundation.